\documentclass[modern]{aastex63}

\usepackage{amsmath}
\usepackage{amssymb}
\usepackage{rotating}

\usepackage{xcolor}
\shorttitle{Starlink Satellites in the ZTF Data}
\shortauthors{P. Mr\'oz et al.}

\begin{document}

\title{Impact of the SpaceX Starlink Satellites on the Zwicky Transient Facility Survey Observations}

\correspondingauthor{Przemek Mr\'oz}
\email{pmroz@astrouw.edu.pl}

\author[0000-0001-7016-1692]{Przemek Mr\'oz}
\affil{Astronomical Observatory, University of Warsaw, Al. Ujazdowskie 4, 00-478 Warszawa, Poland}

\author[0000-0002-9789-2564]{Angel Otarola}
\affil{European Southern Observatory, Alonso de C\'ordova 3107, Vitacura, Regi\'on Metropolitana, Chile}

\author[0000-0002-8850-3627]{Thomas A. Prince}
\affil{Division of Physics, Mathematics and Astronomy, California Institute of Technology, Pasadena, CA 91125, USA}

\author[0000-0002-5884-7867]{Richard Dekany}
\affil{Caltech Optical Observatories, California Institute of Technology, Pasadena, CA  91125, USA}

\author[0000-0001-5060-8733]{Dmitry~A.~Duev}
\affil{Weights \& Biases, Inc, 1479 Folsom St., San Francisco, CA 94103, USA}
\affil{Division of Physics, Mathematics and Astronomy, California Institute of Technology, Pasadena, CA 91125, USA}

\author[0000-0002-3168-0139]{Matthew J. Graham}
\affil{Division of Physics, Mathematics and Astronomy, California Institute of Technology, Pasadena, CA 91125, USA}

\author[0000-0001-5668-3507]{Steven L. Groom}
\affil{IPAC, California Institute of Technology, 1200 E. California Blvd., Pasadena, CA 91125, USA}

\author[0000-0002-8532-9395]{Frank J. Masci}
\affil{IPAC, California Institute of Technology, 1200 E. California Blvd., Pasadena, CA 91125, USA}

\author[0000-0002-7226-0659]{Michael S. Medford}
\affil{University of California, Berkeley, Department of Astronomy, Berkeley, CA 94720, USA}
\affil{Lawrence Berkeley National Laboratory, 1 Cyclotron Rd., Berkeley, CA 94720, USA}

\begin{abstract}
There is a growing concern about an impact of low-Earth-orbit (LEO) satellite constellations on ground-based astronomical observations, in particular, on wide-field surveys in the optical and infrared. The Zwicky Transient Facility (ZTF), thanks to the large field of view of its camera, provides an ideal setup to study the effects of LEO megaconstellations---such as SpaceX's Starlink---on astronomical surveys. Here, we analyze the archival ZTF observations collected between 2019 November and 2021 September and find 5301 satellite streaks that can be attributed to Starlink satellites. We find that the number of affected images is increasing with time as SpaceX deploys more and more satellites. Twilight observations are particularly affected---a fraction of streaked images taken during twilight has increased from less than 0.5\% in late 2019 to 18\% in 2021 August. We estimate that once the size of the Starlink constellation reaches 10,000, essentially all ZTF images taken during twilight may be affected. However, despite the increase in satellite streaks observed during the analyzed period, the current science operations of ZTF are not yet strongly affected. We also find that redesigning Starlink satellites (by installing visors intended to block sunlight from reaching the satellite antennas to prevent reflection) reduces their brightness by a factor of $4.6\pm0.1$ with respect to the original design in $g$, $r$, and $i$ bands.
\end{abstract}


\section{Introduction} \label{sec:intro}

Since 2019 May when SpaceX launched the first batch of their Starlink constellation satellites, the astronomical community has expressed concerns about the constellation's possible impact on astronomical observations. Currently (as of 2021~November~9), the Starlink constellation consists of 1667 satellites\footnote{\url{https://planet4589.org/space/stats/star/starstats.html}} in low-Earth orbit (LEO), and an additional 124 satellites have already reentered Earth's atmosphere due to malfunction or were deliberately retired. Starlink satellites are launched in batches, with up to 60~satellites launched at a time. Fourteen Starlink batches were deployed in 2020 and seventeen are planned to be launched in 2021. In total, SpaceX has received approval by the US Federal Communications Commission (FCC) to operate 12,000 satellites and has submitted filings for 30,000 additional Starlink satellites. Similar LEO satellite constellations are planned by other companies (such as OneWeb, Amazon, Samsung) and national agencies. If these plans were fully put into action, it is estimated that up to 100,000 satellites may be launched into LEO in the upcoming decade \citep[e.g.,][]{hainaut2020,williams2021}, polluting the night sky and potentially impacting astronomical observations.

The first 12,000 Starlink satellites are planned to be arranged in five shells with different operational altitudes ranging from 540 to 570~km \citep{williams2021}; most of the already deployed satellites are at the 550~km orbit. (However, SpaceX has applied to the FCC for a second-generation constellation with considerably lower altitudes from 328 to 614\,km.) Every satellite has three stages of flight: orbit raise, when the satellites are using their thrusters to raise altitude over the course of a few weeks, parking orbit (380~km above Earth) to adjust the orbital plane, and operational orbit (540--570~km above Earth).\footnote{\url{https://www.spacex.com/updates/starlink-update-04-28-2020/index.html}\label{footnote:spacex}} The satellites may be visible to the naked eye during the first two stages of flight, however, this phase is relatively short. Once the satellites reach their operational orbit, they reconfigure so that their antennas face Earth and the visibility of their solar panels from the ground is reduced. As a result, satellites become darker. Satellites may become brighter again during de-orbiting but this end-of-life stage is short.\footnote{While the brightest orbit raise and de-orbiting phases may be brief, if satellite populations are growing and/or large satellite populations are being maintained, these phases become non-negligible.} Because of their low orbital altitudes, the projected surface density of Starlink satellites is greatest near the horizon and during twilight. Simulations by \citet{mcdowell2020} and \citet{hainaut2020} indicate that about 90\% of Starlink satellites in range appear concentrated along the horizon (below $30^{\circ}$ of elevation) and the number of illuminated satellites plummets as the Sun reaches $20^{\circ}-30^{\circ}$ of elevation below the horizon. Thus, it is expected that LEO satellite constellations may particularly affect twilight observations.

Several studies have been undertaken to understand the impact of satellite constellations on astronomical observations. These include the reports summarizing the outcomes of the Satellite Constellations~1 and 2 (SATCON1 and SATCON2) workshops\footnote{\url{https://aas.org/sites/default/files/2020-08/SATCON1-Report.pdf},\\ \url{https://noirlab.edu/science/events/websites/satcon2/publications}} \citep{walker2020,satcon2} and Dark and Quiet Skies for Science and Society workshop\footnote{\url{https://www.iau.org/static/publications/dqskies-book-29-12-20.pdf}} as well as the JASON group report prepared for the US National Science Foundation.\footnote{\url{https://www.nsf.gov/news/special_reports/jasonreportconstellations/}} In particular, \citet{walker2020} list several science areas that are particularly vulnerable to the impacts of large LEO satellite constellations, including wide-field imaging in optical and infrared, searches for near-Earth objects (NEOs), observations of rare transients (e.g., gravitational wave events, gamma-ray bursts, fast radio bursts), or imaging of large, extended low-surface-brightness targets, among others.

To address concerns from the scientific community, SpaceX decided to redesign Starlink satellites to mitigate their impact on astronomical observations. One of the satellites, STARLINK-1130 (``DarkSat''), had an experimental coating to make it less reflective. However, this solution led to thermal issues and was subsequently abandoned.\footnote{See footnote \ref{footnote:spacex}.} More recent satellites (STARLINK-1436 and STARLINK-1522 onward, also called ``VisorSats'') have visors to block sunlight from reflecting from parts of the satellite. We cannot exclude, however, that some of the recent Starlink satellites have not deployed visors, for example, due to technical failure. Visors are intended to reduce the brightness of satellites in operational orbit.

The current studies of the impact of LEO satellites on astronomical observations are based either on simulations \citep{mcdowell2020,hainaut2020,williams2021,bassa2021,lawler2021} or targeted follow-up observations \citep{tregloan2020,tregloan2021,mallama2021b,mallama2021c}. However, we expect that these impacts should be greatest for wide-field sky surveys, such as the Zwicky Transient Facility (ZTF, \citealt{bellm2019,graham2019,masci2019}), Asteroid Terrestrial-impact Last Alert System (ATLAS, \citealt{tonry2011}), All-Sky Automated Survey for Supernovae (ASAS-SN, \citealt{shappee2014}), and---in the future---Vera Rubin Observatory's Legacy Survey of Space and Time (LSST, \citealt{tyson2020}). In this paper, we analyze the archival ZTF data to study the impact of Starlink satellites on ZTF science operations and to assess the effectiveness of mitigation strategies introduced by SpaceX.

\clearpage 

\begin{longrotatetable}
\begin{deluxetable}{lrcrrrrrrrrrr}
\tabletypesize{\footnotesize}
\tablecaption{ZTF Observations of Starlink Satellites (First 15 Entries)\label{tab:starlink}} 
\tablehead{
\colhead{Name} & \colhead{JD} & \colhead{Filter} & \colhead{$m$} & \colhead{$\omega$} & \colhead{$r$} & \colhead{$a_{\odot}$} & \colhead{$A_{\odot}$} & \colhead{$a$} & \colhead{$A$} & \colhead{$\lambda_0$} & \colhead{$\varphi_0$} & \colhead{$H$} \\
\colhead{} & \colhead{} & \colhead{} & \colhead{(mag)} & \colhead{(deg\,s$^{-1}$)} & \colhead{(km)} & \colhead{(deg)} & \colhead{(deg)} & \colhead{(deg)} & \colhead{(deg)} & \colhead{(deg)} & \colhead{(deg)} & \colhead{(km)}}
\startdata
STARLINK-49 & 2458831.60592 & g & $7.04 \pm 0.03$ & 0.258 & $962.438$ & $-22.66$ & 256.35 & 27.70 & 217.33 & $-122.200$ & 26.878 & 530.583 \\
STARLINK-1038 & 2458832.04574 & g & $2.42 \pm 0.03$ & 0.712 & $529.803$ & $-19.70$ & 105.39 & 55.86 & 161.74 & $-117.095$ & 31.816 & 474.684 \\
STARLINK-55 & 2458833.57171 & r & $5.88 \pm 0.05$ & 0.488 & $685.030$ & $-12.67$ & 250.29 & 50.24 & 200.94 & $-119.178$ & 29.052 & 530.968 \\
STARLINK-27 & 2458833.59185 & r & $4.89 \pm 0.07$ & 0.668 & $600.089$ & $-18.43$ & 253.76 & 58.87 & 340.83 & $-119.274$ & 34.571 & 531.714 \\
STARLINK-58 & 2458854.59997 & g & $6.04 \pm 0.03$ & 0.545 & $653.377$ & $-18.38$ & 254.60 & 47.43 & 163.40 & $-116.782$ & 30.604 & 531.064 \\
STARLINK-59 & 2458854.60132 & g & $5.75 \pm 0.05$ & 0.614 & $615.633$ & $-18.77$ & 254.83 & 55.11 & 182.10 & $-117.766$ & 30.996 & 520.999 \\
STARLINK-78 & 2458854.62476 & r & $6.12 \pm 0.02$ & 0.414 & $967.563$ & $-25.63$ & 258.83 & 30.09 & 226.65 & $-122.622$ & 28.418 & 530.365 \\
STARLINK-70 & 2458855.58194 & g & $6.94 \pm 0.02$ & 0.371 & $1033.708$ & $-13.05$ & 251.53 & 29.50 & 145.73 & $-115.430$ & 30.422 & 520.402 \\
STARLINK-1024 & 2458860.58730 & r & $6.24 \pm 0.02$ & 0.502 & $805.491$ & $-13.81$ & 252.84 & 26.75 & 237.73 & $-123.848$ & 30.880 & 397.554 \\
STARLINK-1024 & 2458860.58775 & r & $6.32 \pm 0.03$ & 0.499 & $807.748$ & $-13.94$ & 252.92 & 26.71 & 228.24 & $-121.979$ & 29.058 & 397.300 \\
STARLINK-1059 & 2458860.59268 & r & $7.66 \pm 0.05$ & 0.446 & $920.345$ & $-15.36$ & 253.78 & 22.05 & 230.11 & $-124.099$ & 29.346 & 401.174 \\
STARLINK-1060 & 2458860.59762 & r & $4.84 \pm 0.06$ & 0.394 & $1042.920$ & $-16.79$ & 254.65 & 18.34 & 230.34 & $-125.268$ & 28.768 & 403.261 \\
STARLINK-1041 & 2458862.60284 & r & $6.50 \pm 0.05$ & 0.331 & $1096.476$ & $-17.99$ & 255.76 & 24.15 & 270.48 & $-127.233$ & 32.830 & 551.273 \\
STARLINK-1043 & 2458863.58660 & r & $6.73 \pm 0.03$ & 0.398 & $968.809$ & $-13.12$ & 253.03 & 31.89 & 212.72 & $-121.156$ & 27.741 & 550.205 \\
STARLINK-40 & 2458864.07138 & r & $7.42 \pm 0.04$ & 0.333 & $1192.386$ & $-14.22$ & 106.19 & 22.08 & 136.25 & $-110.999$ & 25.181 & 530.148 \\
\dots & \dots & \dots & \dots & \dots & \dots & \dots & \dots & \dots & \dots & \dots & \dots & \dots \\
\enddata
\tablecomments{This table contains the following information: satellite name, Julian Date of observation, filter, magnitude and magnitude error $m$, mean angular velocity $\omega$, range $r$, elevation of the Sun $a_{\odot}$, azimuth of the Sun $A_{\odot}$, elevation of the trail center $a$, azimuth of the trail center $A$, longitude of the subsatellite point $\lambda_0$, latitude of the subsatellite point $\varphi_0$, and altitude of the satellite $H$.\\(This table is available in its entirety in machine-readable form.)}
\end{deluxetable}
\end{longrotatetable}

\section{Zwicky Transient Facility}

ZTF is a new optical time-domain survey that uses the Palomar 48 inch Schmidt-type Samuel Oschin telescope. The survey started regular science observations on 2018~March~20. The telescope is equipped with a custom-build wide-field camera that provides a 47\,deg$^2$ field of view, one of the largest worldwide. Thanks to the enormous field of view, ZTF can survey the sky at a rate of $\approx3760$\,deg$^2$ per hour to a depth of $\approx\,20.5$ mag. A typical exposure time is 30\,s although sometimes deeper images are taken. 

The ZTF camera consists of sixteen science CCD detectors, each with $6144 \times 6160$ pixels \citep{bellm2019,dekany2020}. Each detector is divided into four amplifier channels (``quadrants''), so each science exposure consists of 64 separate images. Thus, a satellite moving across the field of view of the telescope can be detected on multiple images (quadrants). The plate scale of the camera is $1.01''$ per pixel. Observations are taken through three filters: ZTF-$g$, ZTF-$r$, and ZTF-$i$ (hereafter called $g$, $r$, and $i$, respectively), whose transmission curves are presented by \citet{dekany2020}.

ZTF carries out several surveys \citep[e.g.,][]{bellm_scheduler2019}. Currently, since 2020~October, 50\% of available observing time is allocated to a public survey of the entire Northern sky in $g$ and $r$ bands with a two-night cadence. The remaining time is dedicated to smaller surveys, including the ``Twilight Survey,'' which is designed to observe portions of the sky as close as possible to the Sun during evening and morning twilight to search for solar system objects at small elongations \citep{ip2020}.

\section{Identification of Starlink Trails}

Bright satellite and aircraft streaks are detected and masked on science images by the ZTF data processing system (which is presented in depth by \citealt{masci2019}). Satellite streaks are identified using the \textsc{CreateTrackImage} software that is described in detail by \citet{laher2014}. In short, the algorithm searches for contiguous blobs of pixels that are at or above the local image median plus 1.5 times the local image-data dispersion. The blobs are classified as tracks depending on their length and width. See \citet{laher2014} for more details. Pixels located within the blob are masked and their location is stored in a separate (``mask'') file. The algorithm can identify and mask multiple satellite trails on one image.

To identify Starlink satellite trails on ZTF images, we use Starlink orbital elements encoded in the two-line element sets (TLEs) provided by CelesTrak.\footnote{\url{https://celestrak.com/NORAD/elements/starlink.txt}} These TLEs (also called ``supplemental TLEs'') are derived from latest Starlink ephemeris data provided by SpaceX and allow one to predict the position of the satellite with the precision of 500\,m or better.

We then calculate the expected equatorial coordinates of every satellite for every science image using the \textsc{PyEphem} astronomy library\footnote{\url{https://rhodesmill.org/pyephem/}} and check if the satellite crossed the field of view of the telescope during a given pointing and was illuminated by the Sun at that time. The TLEs that we use and the images that we check span the period from 2019~November~18 through 2021~September~7.

Once we identify images that may be affected by Starlink trails, we use the ``mask'' file to find the center of the streak on every CCD quadrant. We search for streaks within 1~deg of the position predicted by TLEs. We then measure the brightness of a 50~pixel portion of the streak using the trail-fitting technique described by \citet{veres2012} and \citet{ye2019}. Every streak may be detected on multiple quadrants, in which case the reported magnitude is the average (with outlier rejection) of many  individual measurements. The reported magnitude uncertainty is the sum (in quadrature) of the standard error of the mean and 0.02~mag, which is the accuracy of calibrations. Our measurements for 5301 streaks are presented in Table~\ref{tab:starlink}. In addition to measured magnitudes, Table~\ref{tab:starlink} contains the following information: mean angular velocity of the satellite, range, elevation and azimuth of the Sun, elevation and azimuth of the trail center, longitude and latitude of the subsatellite point, and altitude of the satellite.

\begin{figure}
\centering
\includegraphics[width=0.8\textwidth]{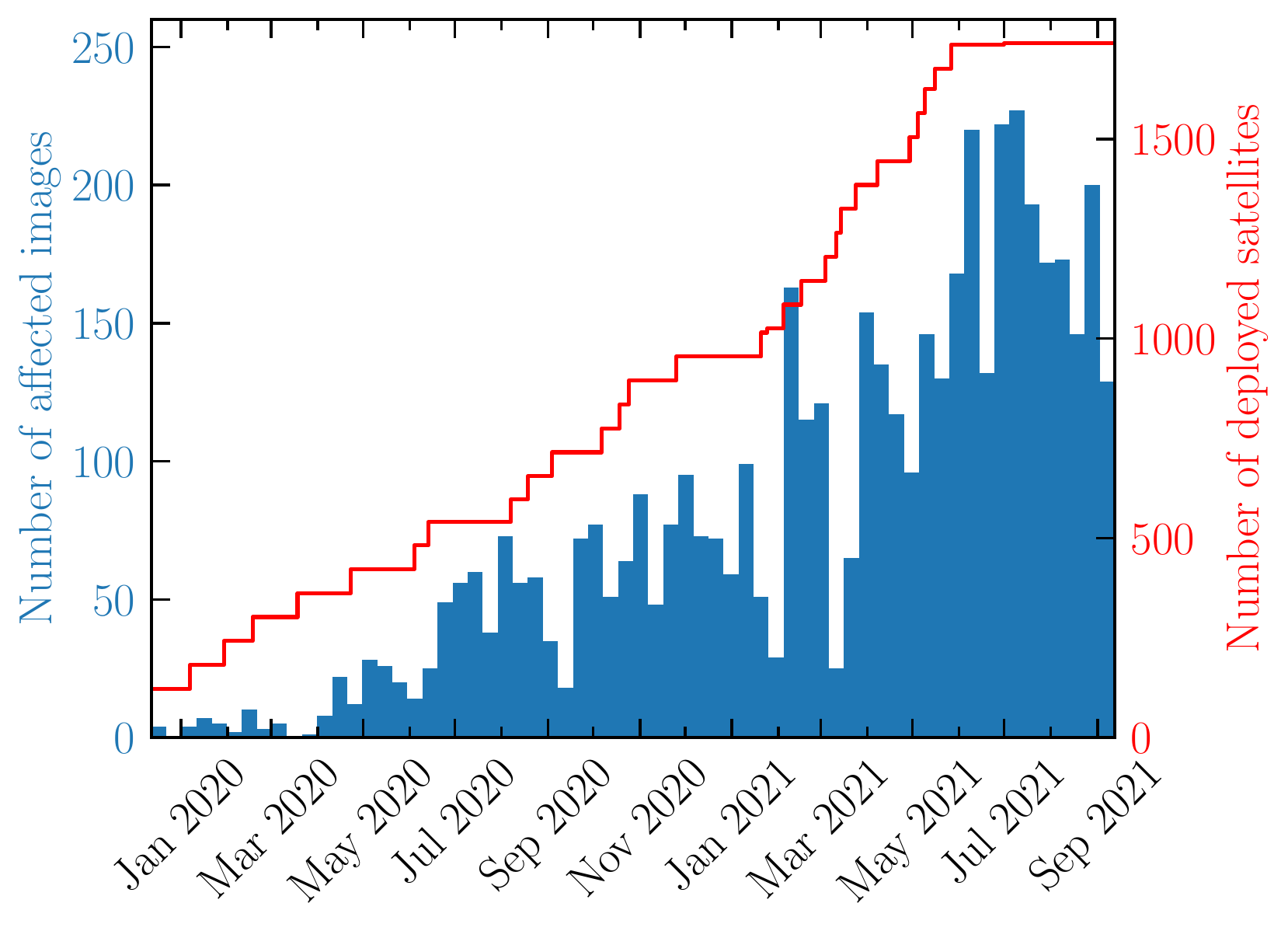}
\caption{Number of ZTF images with at least one Starlink satellite trail in 10~day bins. The thick red line presents the cumulative number of deployed satellites.}
\label{fig:histogram}
\end{figure}

\begin{figure}
\centering
\includegraphics[width=0.8\textwidth]{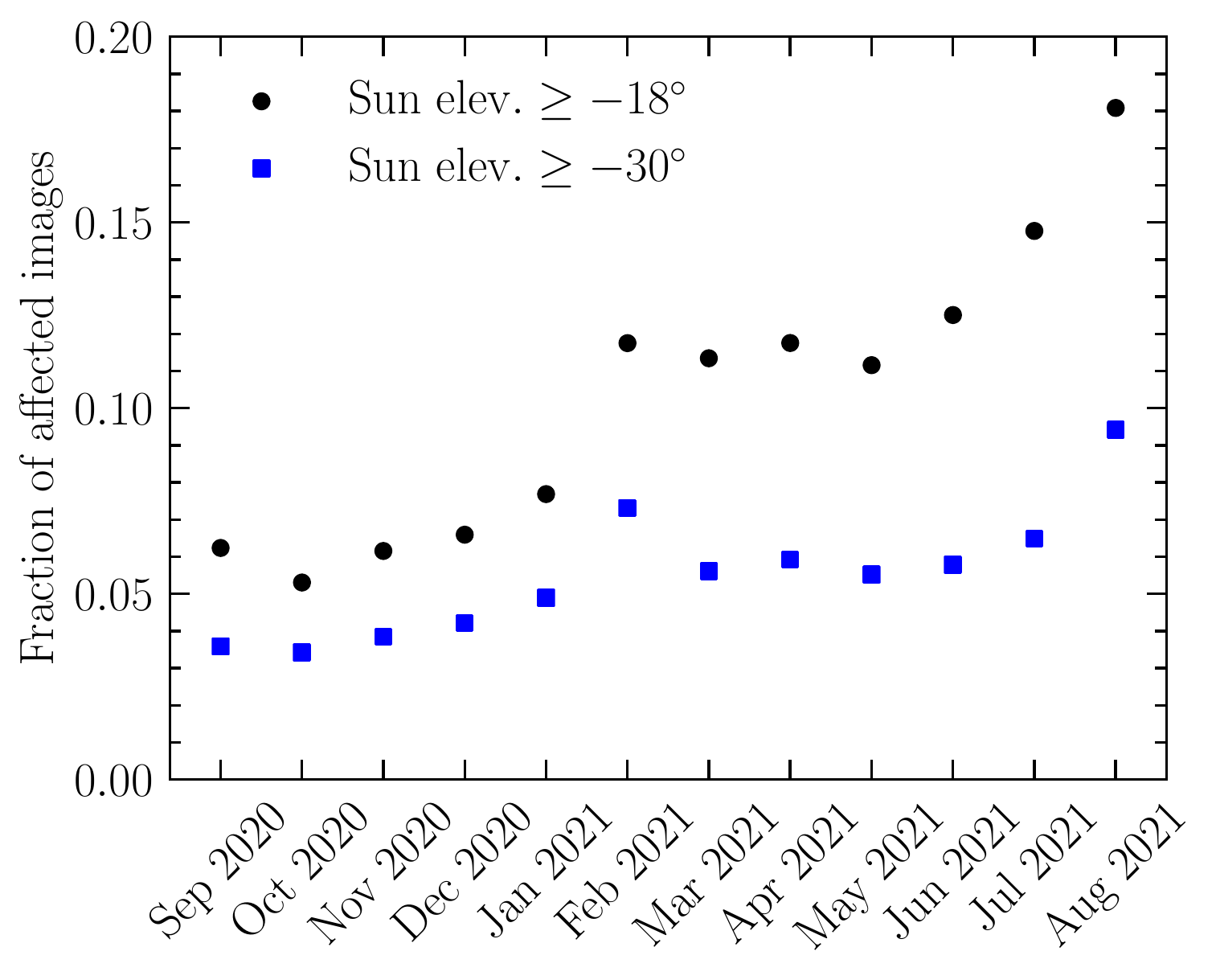}
\caption{Fraction of ZTF images with at least one Starlink satellite trail. Black dots (blue squares) mark images taken when the solar elevation was greater than $-18^{\circ}$ ($-30^{\circ}$). Currently, nearly 20\% of images taken during twilight are affected.}
\label{fig:frac_aff_images}
\end{figure}

\section{Frequency of the Trails and Their Impact on ZTF Science Operations}

Figure~\ref{fig:histogram} presents a number of ZTF images with at least one Starlink satellite trail in 10~day bins. The number of affected images is clearly increasing with time: from less than one affected image per night in early 2020 to nearly 20 streaked images per night in the second half of 2021. Multiple satellites (up to 15) may be visible in one image, and, on average, there are 1.09 satellite streaks per affected image. Figure~\ref{fig:histogram} also shows the cumulative number of deployed satellites from the Starlink constellation. Currently about 1.2\% of all deployed satellites are detected by ZTF in a given night.

As expected by \citet{mcdowell2020} and \citet{hainaut2020}, twilight and high-airmass observations are particularly vulnerable to contamination from LEO satellite trails. About 64\% of the detected trails were imaged during twilight (solar elevation greater than $-18^{\circ}$) and about 70\% (45\%) were at elevation lower than $40^{\circ}$ ($30^{\circ}$). Figure~\ref{fig:frac_aff_images} presents the fraction of ZTF images (taken when the solar elevation was greater than $-18^{\circ}$ and $-30^{\circ}$) with at least one Starlink satellite trail. In late 2020, about 6\% of images taken during twilight were affected, and this fraction went up to $\approx 18\%$ in 2021~August. We expect that essentially all ZTF images taken during twilight may be affected once the size of the constellation exceeds $\approx 10,000$. If the entire constellation of 42,000 was deployed, every image taken during twilight would be contaminated with about four satellite streaks.

Although these numbers seem to be ominous, currently, the overall impact of Starlink satellites on ZTF---as measured by the fraction of pixels that are lost---is not large. 
Assuming that the streaks are distributed randomly, the average streak duration is about 7~deg. The width of the masked area is about $10''$, so the area lost due to a single satellite trail is $\approx 0.02$\,deg$^2$, which is $4\times 10^{-4}=0.04\%$ of the entire detector.

To estimate the fraction of pixels that are lost for the expected constellation of 42,000 Starlink satellites, and knowing that currently about 18\% of all images taken during twilight are affected with about 1600 satellites deployed, we use a simple scaling $0.04\% \times 0.18 \times 42000 / 1600=0.2\%$. For images taken when the solar elevation is greater than $-30^{\circ}$, this fraction is 0.1\%. Taking into account that about 36\% of all ZTF observations are taken when the solar elevation is greater than $-30^{\circ}$ (and almost no trails are visible if the solar elevation is lower), approximately $4\times 10^{-4}$ of all pixels would be lost over the course of a year. However, simply counting pixels affected by satellite streaks does not capture the entirety of the problem, for example resources that are required to identify satellite streaks and mask them out or the chance of missing a first detection of an object (a transient or a solar system object).

Since LEO satellite trails disproportionately affect twilight and high-airmass observations (in particular, the ZTF Twilight Survey), their impact is largest on the solar system science cases, such as the search for and follow-up of comets, asteroids, NEOs, and interstellar asteroids. Satellite trails may also disrupt time-critical observations of rare transient phenomena, such as gravitational wave triggers by LIGO and Virgo, gamma-ray bursts, or neutrino counterparts. However, the probability that the optical transient is missed due to a satellite trail is currently small (less than 0.04\%).

So far, ZTF science operations have not yet been severely affected by satellite streaks, despite the increase in their number observed during the analyzed period. This is thanks to a combination of factors: an efficient image-reduction pipeline, which is able to detect and mask satellite trails, a fact that trails are much fainter than the image saturation limit, and a high survey cadence, among others. We acknowledge, however, that this may not be the case for other wide-field surveys, such as Rubin/LSST \citep{tyson2020}, or projects that are focused on nontransient science (for example, measurements of weak gravitational lensing cosmic shear), which may be affected by systematic effects introduced by satellite trails.

Our results are qualitatively consistent with simulations presented by \citet{hainaut2020}. \citet{hainaut2020} found that 30\% of exposures collected by Rubin/LSST during twilight would be affected due to satellite trails assuming a full constellation of 80,000 LEO satellites and a 15~s exposure time. The field of view of the ZTF camera is five times larger than that of Rubin/LSST and typical exposure times are twice as long. Thus, ZTF should observe on average $0.3\times 5 \times 2 = 3$ satellite trails on every image taken during twilight assuming the model of \citet{hainaut2020}, which is a factor of $\sim2-3$ lower than our observations indicate. However, this difference mainly results from different observing strategies adopted by ZTF and \citet{hainaut2020}, who assumed that all observations are taken at elevations greater than $30^{\circ}$, whereas ZTF is actively observing low-elevation fields close to the Sun during twilight.

\section{Satellite Brightness Distribution}

While the influence of Starlink satellites on ZTF is not large, one of the major concerns is the impact of LEO satellite trails on science operations of Rubin/LSST. Bright trails may be saturated in Rubin/LSST images, causing image artifacts, which may make part of the exposure unusable. Thus, one of the recommendations of the SATCON1 workshop \citep{walker2020} was that the satellites should be fainter than 7~mag in the $V$($g$) band (assuming their range of 550\,km). This limit corresponds to a brightness level that enables artifact correction of Rubin/LSST images to well below background noise \citep{tyson2020}.

In response to the concerns of the astronomical community, SpaceX redesigned their satellites by including visors intended to prevent the sunlight from illuminating the surface of the satellite bus facing normal to the Earth's surface and intended to reduce the sunlight reflected by the satellites, thus making them fainter.

\begin{figure}
\centering
\includegraphics[width=0.49\textwidth]{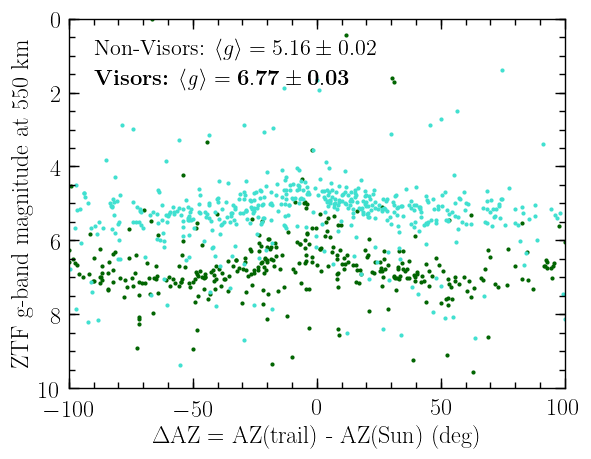}
\includegraphics[width=0.49\textwidth]{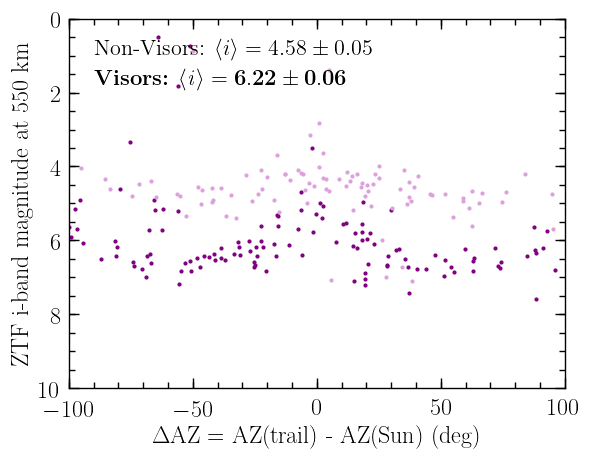}\\
\includegraphics[width=0.49\textwidth]{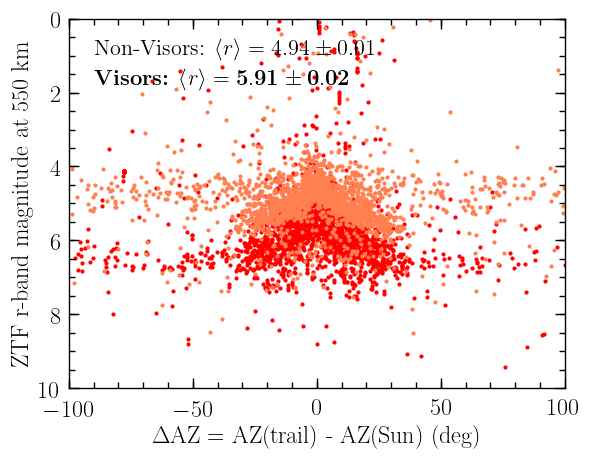}
\includegraphics[width=0.49\textwidth]{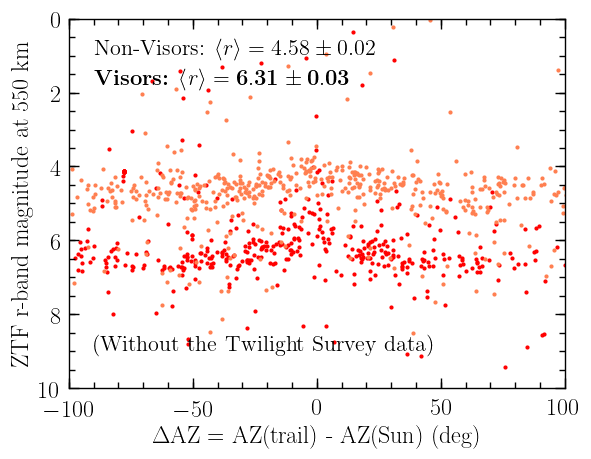}\\
\caption{Starlink streak brightness measurements in three filters used by ZTF, $g$, $r$, and $i$, as a function of the azimuth difference between the center of the trail and the Sun. Magnitudes are scaled to a common range of 550\,km. Darker colors (green, red, purple) mark visor-type satellites, lighter colors (turquoise, orange, plum) mark (non-visor) original design satellites. The mean magnitudes printed in each panel are also reported in Table~\ref{tab:means}.}
\label{fig:magnitudes}
\end{figure}

Figure~\ref{fig:magnitudes} presents our Starlink streak brightness measurements in three filters used by ZTF, $g$, $r$, and $i$, as a function of the azimuth difference between the center of the streak and the Sun. Satellites were located at a different distance (range) from the observer during these measurements (median: 926\,km, 90\% of measurements were taken at range from 562 to 1256\,km), so the magnitudes presented in Figure~\ref{fig:magnitudes} are scaled to a common range of 550\,km (which is the nominal orbital height of the Starlink constellation):
\begin{equation}
m_{550\,\mathrm{km}} = m - 5\log (r/550\,\mathrm{km}),
\end{equation}
where $r$ is the satellite range in km.

Satellites are brightest when their solar elongation is smallest and then their brightness decreases to a minimum when the difference in the azimuth angle is about $70^{\circ}$. For larger angles, the satellite brightness increases again. This effect may be due to the fact that when the satellite and the Sun are at the same azimuth, the brightness of the satellite is the sum of two emission components, a diffuse and a specular reflection of light in the forward direction (across the longer dimension of the Earth-facing surface of the satellite). As the satellite gets further from the Sun, we see mainly diffuse reflected sunlight across the smaller dimension of the satellite bus. A similar dependence of the satellite brightness on the azimuth difference was reported by \citet{mallama2021b}.

The distributions of streak magnitudes in all three filters are evidently bimodal. Table~\ref{tab:means} presents the mean brightness (with outlier rejection) of original and visor-type satellites. Visor-type satellites are, on average, 1.61, 0.97, and 1.64 mag fainter in $g$, $r$, and $i$ bands, respectively. This corresponds to a brightness reduction by a factor of 4.4, 2.4, and 4.5, respectively. It is interesting to note that the brightness reduction in the $r$ band is smaller than in $g$ and $i$ bands. This difference comes from the ZTF Twilight Survey observations which are collected in the $r$ band. 

The Twilight Survey observations are different from typical images taken by ZTF because the target fields are located close to the Sun (solar elongation smaller than $60^{\circ}$) and the Sun is located just below the horizon (elevation larger than $-20^{\circ}$). 
The original design (non-visor) satellites observed as a part of the Twilight Survey were, on average, 0.47~mag fainter than during other $r$-band observations (orange dots in the lower-left panel of Figure~\ref{fig:magnitudes}). The satellites were darker than they would be in the non-twilight data because of either a shadow or a specular reflection (not pointing to the observer). For visor-type satellites (red dots in the lower-left panel of Figure~\ref{fig:magnitudes}), this behavior is no longer present. The net effect is that the brightness difference between visor and non-visor satellites is smaller for the Twilight Survey data ($0.77\pm0.03$~mag in the $r$-band). If the Twilight Survey observations (Sun elevation larger than $-20^{\circ}$ and solar elongation smaller than $60^{\circ}$) are removed, the mean magnitude difference in the $r$ band is 1.73~mag (brightness reduction by a factor of 4.9), which is consistent with measurements collected in other bands.

The mean brightness reduction in all three bands (without the Twilight Survey data) is $1.67 \pm 0.03$ mag. Thus, the use of visors reduces the brightness of Starlink satellites by a factor of $4.6 \pm 0.1$ irrespective of the filter used. The brightness reduction that we measure is larger than that reported by \citet{mallama2021c} (1.32~mag).

The mean colors of visor-type satellites are $g-r=0.46\pm0.05$~mag and $r-i=0.09\pm0.07$~mag, and are consistent with the solar colors ($g-r=0.45$~mag and $r-i=0.12$~mag) \citep{holmberg2006}. Visor-type satellites reflect the sunlight neutrally and their magnitudes in different optical filters can be estimated using solar colors.

Finally, we also find that of the 377 normalized apparent magnitudes of visor-type satellites in the $g$ band, 253 (that is, 67\%) are still brighter than the Recommendation \#5 of the SATCON1 Workshop ($g=7$ mag). Thus, although mitigation strategies adopted by SpaceX lead to substantial darkening of the satellites, they are not sufficient to reach the goal set during the SATCON1 workshop.

\begin{table}
\caption{Mean brightness (at 550\,km) of original and visor-type Starlink satellites in ZTF $g$, $r$, and $i$ bands.}
\label{tab:means}
\begin{tabular}{lccc}
\hline \hline
Filter & Non-Visors & Visors & Difference \\
\hline
$g$ & $5.16 \pm 0.02$ & $6.77 \pm 0.03$ & $1.61 \pm 0.04$ \\
$r$ (all) & $4.94 \pm 0.01$ & $5.91 \pm 0.02$ & $0.97 \pm 0.03$ \\
$r$ (without Twilight Survey data) & $4.58 \pm 0.02$ & $6.31 \pm 0.03$ & $1.73 \pm 0.04$ \\
$r$ (only Twilight Survey data) & $5.05 \pm 0.01$ & $5.82 \pm 0.02$ & $0.77 \pm 0.03$ \\
$i$ & $4.58 \pm 0.05$ & $6.22 \pm 0.06$ & $1.64 \pm 0.08$ \\
\hline
\end{tabular}
\end{table}

\section{Summary}

Archival ZTF data, thanks to the large field of view of the ZTF camera and the survey speed, allow us to study the effects of LEO megaconstellations (such as Starlink) on scientific operations of ground-based survey telescopes. They are of particular interest because they are archival, not targeted, and therefore represent what typical large-area surveys are seeing. In this Letter, we analyze the archival ZTF observations of Starlink satellites collected between 2019~November and 2021~September. We find 5301 satellite streaks that can be attributed to Starlink satellites.

The number of images affected by satellite trails is alarmingly growing as more and more Starlink satellites are being deployed in orbit. Twilight observations are particularly affected---a fraction of streaked images taken during twilight has increased from 6\% in late 2020 to 18\% in 2021~August. We estimate that once the size of the Starlink constellation reaches 10,000, essentially all ZTF images taken during twilight may be affected. However, to date, ZTF science operations have not been severely affected by satellite streaks, despite the increase in their number observed during the analyzed period.

Our data clearly show that visor-type satellites are dimmer. Their brightness is reduced by a factor of $4.6\pm0.1$ with respect to the original design satellites in $g$, $r$, and $i$ bands, which confirms that the mitigation strategies adopted by SpaceX lead to substantial darkening of the satellites. However, many of our observations of visor-type satellites ($\approx~67\%$) are still brighter than the recommendation \#5 of the SATCON1 workshop.

ZTF continues to monitor the Northern sky and we expect it will record thousands of LEO satellite streaks more. New ZTF observations will allow us to monitor the problem and to provide input to the astronomical community as well as the satellite operators.

\section*{Acknowledgements}

We thank Olivier Hainaut and Radek Poleski for discussions and their comments on the manuscript. We thank Dr.~T.~S.~Kelso of CelesTrak for maintaining the database of Starlink TLE data.
Based on observations obtained with the Samuel Oschin Telescope 48-inch and the 60-inch Telescope at the Palomar Observatory as part of the Zwicky Transient Facility project. ZTF is supported by the National Science Foundation under grant No. AST-2034437 and a collaboration including Caltech, IPAC, the Weizmann Institute for Science, the Oskar Klein Center at Stockholm University, the University of Maryland, Deutsches Elektronen-Synchrotron and Humboldt University, the TANGO Consortium of Taiwan, the University of Wisconsin at Milwaukee, Trinity College Dublin, Lawrence Livermore National Laboratories, IN2P3, University of Warwick, Ruhr University Bochum and Northwestern University. Operations are conducted by COO, IPAC, and UW.

\bibliographystyle{aasjournal}

\end{document}